# DEMONSTRATING MINIATURISED, ENTANGLED PHOTON-PAIR SOURCES ON BOARD NANO SATELLITES TO ENABLE FUTURE QKD MISSIONS

**Rakhitha C.M.R.B. Chandrasekara[a]\*, Robert Bedington[a], Xueliang Bai[a], Karthik Ilangovan[a], Yau Yong Sean[a], Denis Naughton[b], Simon Barraclough[b], Douglas Griffin[b], Russell Boyce[b], Alexander Ling[a,c]**

[a] *Centre for Quantum Technologies, National University of Singapore, Block S 15, 3 Science Drive 2, Singapore, 117543*, rakhitha@u.nus.edu
[b] *University of New South Wales Canberra, School of Engineering and Information Technology, Canberra, ACT, Australia*
[c] *Department of Physics, 2 Science Drive 3 Blk S12, Level 2, Singapore 117551, Singapore*,
\* Corresponding Author

**Abstract**
We report on our progress developing highly-miniaturised, polarisation-entangled, photon pair sources for CubeSats. We have a correlated photon-pair source in orbit in the NUS Galassia 2U CubeSat. We also have an entangled photon pair source in production and a high brightness (1Mcps) entangled photon-pair source in development for our upcoming satellite missions. All our sources are proof-of-principle demonstrations that the hardware necessary for entanglement-based QKD can be miniaturised and made sufficiently robust for operation in nano-satellites. The photon pairs they produce are measured with liquid crystal-based Bell state analysers and Geiger-mode avalanche photo-diodes within the source. These space missions allow our on-the-ground radiation, thermal and vibration tests to be validated and the real-world operation and aging of the source in space to be studied. A BBM92 QKD-capable design of the source has been used in a phase A study of a satellite-to-satellite QKD demonstration mission by the University of New South Wales, Canberra. This mission study uses two 6U CubeSats in LEO and aims to demonstrate QKD over separations of increasing distances as the two CubeSats drift apart.

**Acronyms/Abbreviations**

Centre for Quantum Technologies (CQT)
Commercial Off-The-Shelf (COTS)
National University of Singapore (NUS)
Quantum Key Distribution (QKD)
Low Earth Orbit (LEO)
Technology Readiness Level (TRL)
Size, Weight and Power (SWaP)
Spontaneous Parametric Down-Conversion (SPDC)
Beta Barium Borate (BBO)
Computer Aided Design (CAD)
Liquid Crystal Polarisation Rotator (LCPR)
Geiger Mode - Avalanche PhotoDiode (GM-APD)
Coefficient of Thermal Expansion (CTE)
Small Photon Entangling Quantum System (SPEQS)
University of New South Wales (UNSW)
Quantum Bit Error Rates (QBER).
Pointing, Acquisition & Tracking Subsystem (PATS)

## 1. Introduction

Quantum key distribution (QKD) [1] is a family of cryptographic methods for securely establishing symmetric [2], one-time-pad [3] encryption keys between two distant parties. Unlike many conventional public-key methods [4,5] it is not based on solving hard mathematical functions, but on the quantum states of photons. This is attractive because it provides strong assurances that messages encrypted with these keys today will not be vulnerable to cryptanalysis by more powerful computers in the future [6,7].

There are many implementations of QKD, but the most secure versions use pairs of entangled photons [8]. In these methods one photon in each pair is sent to the two parties who make measurements on their quantum states. To verify the security of the distribution the two parties can perform statistical tests and discuss the results openly to determine if the photons they received were entangled. If the entanglement is below a calculated threshold, they should discard those photons and not use them for encryption keys. If the entanglement is adequate, they can be certain that no evesdropper has sufficient knowledge of the photons to determine the key. They can then use their knowledge of the quantum states – information that has not been publicly shared – as a common source of randomness from which to derive a symmetric encryption key known only to them.

Compared with conventional public-key encryption methods however, QKD is more challenging to implement because individual photons must be shared and correlated between the two parties. Through optical fibres [9] or through ground level free-space links [10] individual photons can travel at most a few hundred kilometres before losses become too high and a trusted node repeater station is required. The more stations there are the more vulnerabilities there are in the network. By locating trusted nodes in space they are much harder to infiltrate and far fewer of them are required. In principle a single, trusted node, LEO satellite can be used to share quantum keys between any two optical ground stations on the planet [11], and this has been demonstrated recently by the Chinese Micius satellite [12].

The Micius satellite is 631kg. It is our goal to develop QKD systems that can be used on much smaller space platforms. CubeSats have already been built for conventional laser communications [13] and many groups are researching high precision PATS on the CubeSat platform. Some of these platforms, e.g. [14,15,16], already in principle satisfy the requirements (~arcmin level body pointing and ~urad level pointing accuracy) to implement the QKD link.

We are focussing on developing rugged, miniaturised, sources of entangled photon pairs that could be combined with such systems to enable QKD links. To achieve this, first we have built a source of correlated photon pairs which we are now testing in-orbit while we prepare an upgraded, entangled version of it for in-orbit testing in 2018.

**2. Miniaturised, correlated photon pair sources**

As an intermediate development towards an entangled photon source we designed a source of SPDC [17] photons that were correlated, but not entangled in polarisation (though they may have entanglement in other degrees of freedom, e.g. time-frequency). This allows us to test optical alignment and gluing techniques, optoelectronics and miniaturised mechanical components that are required for the more complex, entangled source design.

A wavelength stabilised laser diode was chosen at 405 +/- 0.1 nm to produce down converted photons at 760 and 867 nm in a BBO crystal in Type-I geometry. For polarisation analysers, a non-inertial polarisation rotator based on liquid crystals (LCPRs) [18] and a polarisation beam splitter based design has been adopted. The polarisation rotation induced by a LCPR is electrically controllable and such an optoelectronics device effectively removes possible inertial interference with satellite attitude control systems when rotor based solutions are used. This allows the quantum payload to be operated independent of the CubeSat attitude control system.

GM-APDs have been used to detect single photons after polarisation analysers. A new, low power control circuit has been designed and tested to compensate for photon detection efficiency fluctuations of GM-APDs due to temperature variations. In addition to that, the circuit has also extended the linearity of the detector when operated in passively quenched circuits with high input photon rates [19].

The correlated photon pair source is characterised in a thermal chamber over the simulated orbital temperature range of 10-30°C. Subsequently, it was successfully demonstrated in a 35.5 km altitude balloon test [20]. The temperature gradients and low pressure experienced by the correlated photon source was an approximation to space conditions and successful operation throughout the journey lifted the confidence level of the TRL of the payload.

Subsequently in 2016, an improved correlated photon source was successfully demonstrated [21] on board the NUS Galassia 2U CubeSat [22] in 550km LEO for a period of nine months and counting [23]. Figure 1 illustrates the photon correlations observed in space for a span of 261 days since launch and the baseline established on Earth (black). Such a correlation curve is obtained by fixing the polarisation setting of one arm and scanning the polarisation setting of the other arm.

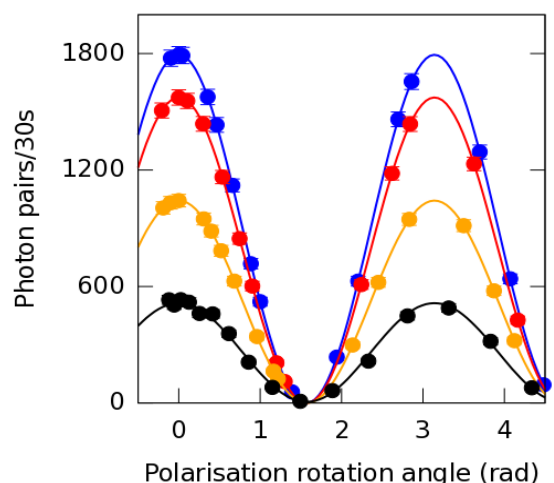

Figure 1. Polarisation correlation baseline curve established on Earth (at 24.7°C) is plotted in black. The three correlation curves plotted in orange, red and blue correspond to orbital days 36, 128 and 261 since launch and the data was collected at 22.1°C, 15°C and 15°C respectively. The peaks and trough correlations counts correspond to similar and orthogonal polarisation settings of the LCPRs respectively.

It is noted that the brightness (i.e. peak correlations in Fig.1) of the source was higher at lower temperatures due to thermal response of the mechanical flexure stage which held the non-linear crystal. This was confirmed by analysing the response of the flexure stage of the backup payload on the ground. At a similar temperature (15°C) the quantum correlations observed over time (day 128 and 261) were similar and prove the robustness of the source.

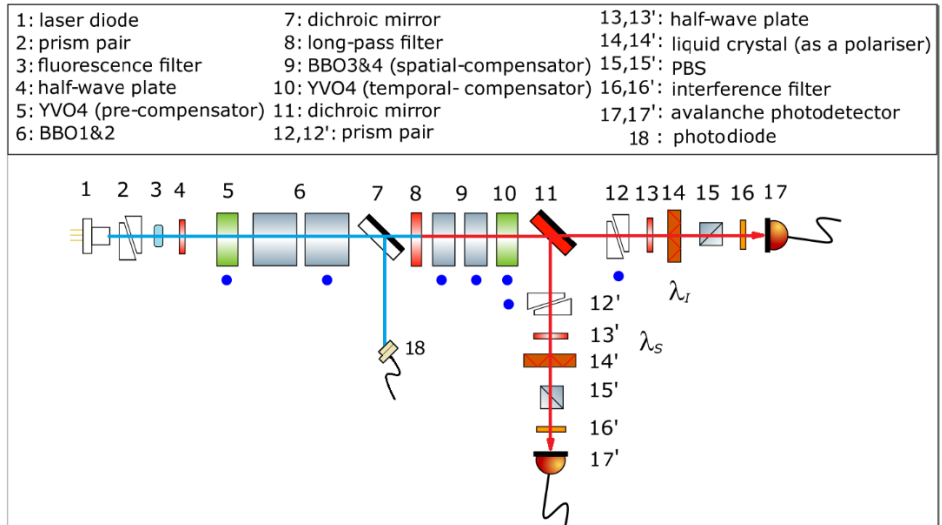

Figure 2. Optical layout of the SPEQS-1 entangled photon pair source for SpooQy-1 CubeSat. Blue dots are additional components compared to the correlated source

A temperature dependent response of the LCPR has also been identified according to the data collected from Galassia. It caused the correlation curve plotted in Figure. 1 to shift right (left) at low (high) temperatures (with respect to the calibrated temperature) when plotted against the applied voltage.

### 3. Miniaturised, entangled photon pair sources

Building on the correlated source heritage, the first generation polarisation-entangled photon pair source (SPEQS-1) is currently under construction. Extra non-linear crystals are added to achieve entanglement resulting in a larger design that requires a 3U CubeSat. The optical design of SPEQS-1 is shown in Figure 1.

Closer attention has been paid to thermal expansion and mechanical creep resulting in titanium flexure stages on a titanium optical bench structure. An isostatic mounting for the payload (shown in Figure 3) has been designed by our collaborators in UNSW Canberra, Australia to isolate the payload from the aluminium spacecraft structure. The temperature dependent response of LCPR identified in the Galassia mission is compensated by a novel, low power capacitance tracking based control system for the LCPR operating in the temperature range of 10-30°C [24].

SpooQy-1 is being assembled from GomSpace components in-house at CQT and will validate the SPEQS-1 performance in the space environment. It is currently scheduled for launch in Q3 2018. A CAD model of the SpooQy-1 satellite is shown in Figure 3.

For space to ground QKD, link losses of ~30 dB are typical [25] and the photon pair production rate (brightness) of SPEQS-1 would not be sufficient for practical applications. Accordingly, work is ongoing on SPEQS-2, an improved source which can produce ~1Mcps polarisation entangled photons [26].

### 4. Future missions

Various options are being investigated for using SPEQS-2 to demonstrate QKD from CubeSats and other platforms. As well as space to ground demonstrations [27], intersatellite QKD is also being investigated. A feasibility study on demonstrating a BBM92 QKD link [28] between two 6U CubeSats in LEO is being conducted by UNSW Canberra and CQT, NUS.

To successfully establish the link, a miniaturised high-precision Pointing And Tracking System (PATS) must provide a polarisation preserving optical link with sufficient optical throughput between the two satellites. Together with satellite body pointing, a laser beacon and two axis fine steering mirrors are baselined to maintain pointing errors less than 5 urads rms between the platforms over a distance of 100 km to allow reliable satellite to satellite QKD.

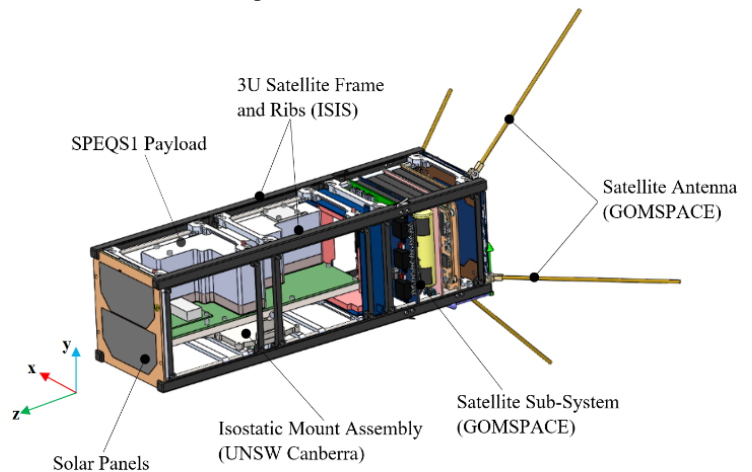

Figure 3. Overview of satellite, with side solar panels, harnesses and interstage panels removed.

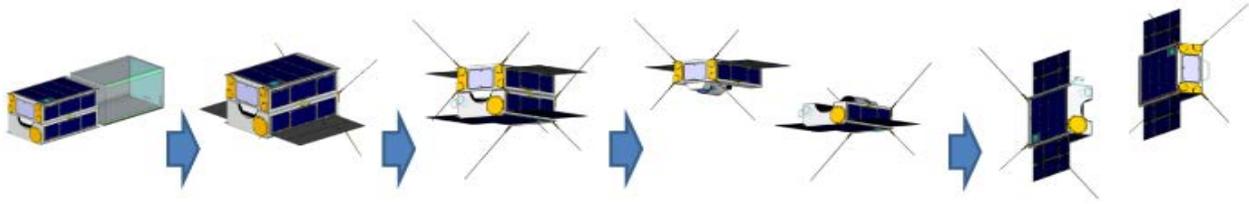

Figure 4. Mission concept of satellite to satellite QKD with two 6U CubeSats.

Figure 4 illustrates the mission concept. Each 6U CubeSat (2x3 configuration) would carry a SPEQS source and detector setup, so both satellites would be capable of operating as a transmitter or a receiver. The two satellites would be launched as a single 12U CubeSat as illustrated in Fig.4 to a sun-synchronous LEO orbit (500-550 km) [29]. The optical telescopes would be co-aligned in the 12U configuration to allow a functionality check of the QKD link prior to separation. After the launch, the CubeSats would separate and manoeuvre to their operational orientations. Once a relative velocity of ~10cm/s is achieved between the spacecraft QKD demonstrations would commence.

**Acknowledgements**

This work is partially supported by the National Research Foundation, Prime Minister's Office, Singapore (under the Research Centres of Excellence programme and through Award No. NRF-CRP12-2013-02) and by the Singapore Ministry of Education.